\documentclass{mn2e}
\begin{document}

\title[Heliopause Structure ]{The Structure Of The Heliopause}

\author[J J Quenby and W R Webber]{J. J. Quenby, $^{1}$\thanks{E-mail: 
j.quenby@imperial.ac.uk} W. R. Webber,$^{2}$\\
$^{1}$Blackett Laboratory, Imperial College,
London, SW7~2BZ, U.K.\\
$^{2}$Department of Astronomy, New Mexico State University
 Las Cruces, NM, 
88003 USA}

\maketitle
\begin{abstract}

Voyager 1 has explored the solar wind-interstellar medium interaction region
between the Terminal Shock and Heliopause, following the intensity distribution
of the shock accelerated anomalous component of cosmic rays in the MeV energy
range. The sudden disappearance of this component at 121.7 AU from the sun is 
discussed in terms of three models for the transition into the interstellar 
plasma flow. Particles trapped flowing parallel to the boundary may penetrate
up to one Larmour radius beyond. If the boundary is stationary, Voyager 1 
directly samples this distance. The boundary could flap, depending on
Heliosheath pressure changes and Voyager 1 then samples the extent of this
motion. Finally, a turbulent bounday layer is considered in which the MeV
particle distribution falls off with distance thus measuring diffusion within 
the layer.  \\

\end{abstract}

\section{INTRODUCTION}

Models and observations related to the interaction of the interstellar medium
(ISM) with the heliosphere identify a solar wind terminal shock 
and a Heliosheath, perhaps 
comprising a low latitude region where the magnetic structure is determined 
by reconnection and a high latitude region where field lines connect back to
the solar wind (Opher el al. 2012). No interstellar bow shock is expected 
(McComas et al. 2012) but the external field pressure causes asymmetry in
the terminal shock (Opher et al. 2006) and may also explain
 intermitent observation
 of shock accelerated cosmic rays (Jokipii et al. 2004, Stone et al. 2005)
 prior to the termination shock crossing.
 Since early models of heliosheath modulation (Potgieter and le Roux 1989, 
Quenby et al. 1990), it has been assumed that the
 Heliopause represented the boundary beyond which the interstellar cosmic ray
intensity would be encountered ( see review by Potgieter (2008)). 
Strauss et al. (2013)
 have recently questioned this assumption and mention
 various possibilities of increased particle scattering as the interstellar
 field wraps around the Heliopause.
The propagation of the anomalous
component, believed to be accelerated at the terminal shock, across the 
heliosheath and into the interstellar medium has been discused by Scherer 
et al. ( 2008) \\
The dramatic Voyager 1 observation of the sudden and sustained disappearance 
of the anomalous component at 121.7 AU (Webber and McDonald, 2013a)
 affords an 
opportunity to explore the properties of the Heliopause boundary. It is the 
purpose of this note to provide some simple alternative models for the
structure of this boundary especially in terms of it's effect on low energy 
anomalous component cosmic rays. Either the boundary is stationary during
 Voyager 1 passage, or it
 flaps significantly and in addition there is possibly a finite width  
particle diffusing layer.\\   
         
\section{VOYAGER DATA}

The starting point in this note lies in data obtained by Webber and McDonald
 (2013a)
from the Voyager 1 CRS instrument (Stone et al. 1977). We concentrate on the 
period of the final drop of the 2-10 MeV proton intensity to $\sim 0.2-0.3$ \%  
of the values seen previously in the Heliosheath. The drop started on 2012.65
and lasted for about 8 weeks (weeks 33-41 of the year) with the spacecraft 
moving 0.07 AU per week. Because there had been two previous intensity drops,
Webber and McDonald (2013a)
 suggest an interpretation in terms of a pulsating boundary
repeatedly crossing Voyager 1. A previous peak in galactic cosmic ray 
intensity suggests an effective region of enhanced modulation at the 
Heliopause. Beyond the drop, the anomalous component is thought to have 
disappeared and intensities close to true interstellar conditions reached. 
At the lowest energy detected, the $>0.5$ MeV channel, the fastest rate of 
decrease represents an e-folding distance $\sim$ 0.01 AU assuming a 
stationary boundary. However, since
 differentiation between various models is helped by looking at the rigidity
dependence of the effect, the decrease is studied between weeks 34 and 35 of
 2012 where the final drop-off can be followed up to 40 MeV in a proton 
channel. Table 1 uses the data from figure 2, Webber and McDonald (2013a)
 to find the observed fractional intensity change $\Delta N/N$
per 0.1 AU at mean energies of
6 MeV and 25 MeV, using data from weeks 41-45 as an interstellar background
level. A stationary boundary is assumed.\\  

\begin{table}
\caption{Stationary Model Boundary Gradients }
\label{symbols}
\begin{tabular}{@{}ccc}
\hline
Proton & $\Delta N/N$     & Larmor Radius   \\
Energy  & per 0.1 AU &       AU               \\
\hline
 6 MeV   & 0.73          & 0.006             \\
25 MeV   & 0.72           & 0.016           \\
\hline
\end{tabular} 
\end{table}

\section{DISCUSSION OF BOUNDARY MODELS}

The published Opher et al. (2006) hydro-magnetic model for the draping of the 
interstellar magnetic field over the Heliosphere shows a $30^{\circ}$ angle 
between field and heliopause at the Voyager trajectory some 10 AU or more 
from this boundary, as seen projected in the meridian plane. 
 However by analogy with the Magnetosphere, a tangential
discontinuity between ISM flow and Heliosheath flow, which has been deflected
along the boundary, is possible. (See also Gurnett et al. (2006))
 In this case, magnetic fields parallel to the
boundary on both sides can occur. In the ISM field B is pointing
in the negative y and z directions where z is solar rotational axis, x in
the interstellar velocity direction and y completing the right handed 
coordinate system. Since the Heliosheath region field is mainly azimuthal  
and inward pointing in the Northern hemisphere, a tangential discontinuity
is likely. If, however, there are fluctuations causing oppositely directed
field components, either side,
as is suggested by the outward pointing boundary
heliosheath field reported by  
Burlaga et al. (2013), reconnection may take place allowing the post
 boundary field to be at some angle $\alpha$ to the boundary. Other ways in
which field lines cross the boundary are if waves described by rotational 
discontinuities are present each side or if the boundary is simply a contact
discontinuity where the plasmas on each side have no relative motion. 
Hence there are several possibilities for the lack of large changes in the
magnetic field vector as Voyager crosses the Heliopause.
Clearly the 
anomalous component protons are following field lines  
and if there is a tangential discontinuty, they can run along the 
boundary just penetrating one Larmor radius, $\rho$, into the ISM. Hence the
radial distribution at the boundary would reflect the pitch angle distribution,
presumably isotropic within the Heliosheath. If however field lines cross the
boundary, there will be a forward cone where particles immediately escape
into the ISM resulting in a
small observable intensity and a region of pitch angle space where 
re-entry into the Heliosheath occurs and the penetration region of pitch
dependent intensity is limited to $\rho cos \alpha$. Using the 
Opher et al. (2012)
  field model value at the Voyager 1 crossing point of $3.8 \mu$G, Table 1
lists the ISM values of $\rho$ at the two energies considered.
 Note
the recently published Voyager 1 field values just beyond the apparent
crossing point are close to our assumed value although there is doubt as 
to whether they represent an extension of local heliosheath conditions
or those farther out in the interstellar medium flow (Burlaga et al. 
2013). 
 The expected
increase of penetration distance with energy predicted by
the stationary boundary model is not borne out by the
data, neither is the observed gradient large enough to match the estimated 
Larmor radii which set the distance scale. Anisotropy data relating to a pitch 
angle distribution is not presented in
 Webber and McDonald (2013a),
but anisotropy is  seen in an equivalent energy range by the LEPC
 instrument,(Krimigis et al., 2013)\\
The observation of three apparent crossings into the ISM suggests boundary 
motion should be considered when evaluating the anomalous component gradient.
Pressure balance across the Heliopause must depend mainly on the near 
isotropic Heliosheath
plasma pressure due to the low energy particles and magnetic field, rather than
the ram pressure of the wind which according to modelling and some
observation, (Krimigis et al. 2011),  
is directed nearly parallel to the boundary. Voyager 2
heliosheath data yield daily changes in total pressure of up to 50\%
(Burlaga et al. 2009). The information that a particular Heliopause equilibrium
 position can no longer be maintained by the isotropic plasma pressure 
can only propagate at the speed of the fastest
wave mode in the subsonic region. Using the post-shock Mach number given in
 the Heliosheath by Borovikov et al. (2011) on the Voyager 1 trajectory,
M=0.58, we obtain a fast mode speed of 228 km/s for the possible collapse 
rate of the boundary. Table 2 then re-estimates the anomalous component 
gradients assuming the boundary scans across the effectively stationary
spacecraft.
\begin{table}
\caption{Moving Model Boundary Gradients }
\label{symbols}
\begin{tabular}{@{}ccc}
\hline
Proton & $\Delta N/N$     & Larmor Radius   \\
Energy  & per  AU &       AU               \\
\hline
 6 MeV   & 0.55          & 0.006             \\
25 MeV   & 0.59           & 0.016           \\
\hline
\end{tabular}
\end{table}
Now the observed region of significant intensity change extends over
$\sim$ 1 AU.  
Hence the anomalous component disappearance could be due to boundary jitter
over 1 AU with an intrinsic boundary
width of 0.01 AU, governed by Larmor motion of the 
Heliosheath trapped particles.\\
Until unambiguous field and plasma flow data become available, the exact 
location and nature of the Heliopause is open to speculation. Here we now 
explore the possibilty of a turbulent, particle diffusing region corresponding
 to the extent of the dramatic 1-40 MeV intensity reduction observed by V1.
 If the boundary
moves it may be expected to drive HM waves into the nearby ISM with properties
initially similar to those of the Heliosheath, before amplitude reduction due 
to divergence into the very large medium sets in.\\
To estimate possible values 
of the diffusion coefficients parallel and perpendicular to the mean field
we will use a theoretical formulation which has achieved reasonable agreement
 with experimental results in the inner Heliosphere, but employing field data
obtained far out in the Heliosheath. These estimates are checked for 
consistency with modelling of low energy cosmic ray modulation in the 
Heliosheath. The waves in the field model are composed 80\% of a 2-dimensional
 component with fluctuation vectors perpendicular to both the mean field
and wave propagation direction and 20\% of a slab component with fluctuations
 perpendicular to the mean field but with wave propagation along the mean field.
As employed by Pei et al. (2010), this composite field model yields a parallel
diffusion mean free path
\begin{equation}
\lambda_{||}=3.13271 \frac{B^{2}}{b_{x,slab}^{2}}(\frac{P}{cB})^{1/3}
\lambda_{slab}^{2/3}\times F
\end{equation}
where $P$ is particle rigidity, $B$ is the mean field, $b_{x,slab}$ is the 
slab component of field fluctuations, $\lambda_{slab}$ is the correlation 
length of the fluctuatiins, assumed valid for all components and F is function
very close to unity in the present application. Based upon a similar theoretical
formulation, Nkosi et al, (2011) find that low energy electron modulation data
are reasonably fitted by a perpendicular diffusion coefficient
\begin{equation}
K_{\perp}=a(\frac{\delta B}{B})^{2}K_{||}
\end{equation}
where the constant $a=0.02$, $\delta B$ is now the total power in
 fluctuations 
and $K_{||}$ is the parallel diffusion coefficient obtained from the mean free
path of equation (1).\\
The most relevant field data is that obtained by Burlaga and Ness
 (2010) in the Heliosheath at 110 AU where the
 field standard deviation is 0.051 nT for a mean
field of 0.08 nT. Bearing in mind the assumption that most flutuation power is 
in a transverse component, we estimate
\begin{equation}
\frac{sd}{B^{2}} \approx \frac{\delta B^{4}}{4B^{4}}
\end{equation}
where $sd$ is the standard deviation of the one day Voyager 1 field
 magnitudes at 110 AU. An estimate for $\lambda_{slab}$ may be obtained from
the break in the power spectrum given by Burlaga and Ness (2010) at 
$2\times10^{-7}$ Hz assuming the relative Voyager 1 plasma speed of 
8 km s$^{-1}$ (Burlaga and Ness 2012). This approach yields 
$\lambda_{slab}=0.27$ AU.
Since the relative speed is 
uncertain, a second estimate comes from assigning the correlation
length to the possible sector structure seen by Burlaga and Ness (2010),
lasting for about 125 days when convected past Voyager 1. Their value would
yield 1.4 AU. Hence using equations (1) and (2) we find Heliosheath values 
for 6 MeV protons of 
$\lambda_{||} \approx 1\rightarrow3.1$ AU and $\lambda_{\perp}\approx 0.04
\rightarrow0.124$ AU. Confirmation of this formulation comes from the work of
Webber et al. (2013b) who fit the observed cosmic ray modulation beyond 105 AU
with a spherically symmetric model approxiation. The radial mean free path,
 which is clearly dominated by perpendicular diffusion, is 0.11 AU at 6 MeV for 
protons.

Anomalous protons escaping from the Heliopause due to random drift and 
resonance scattering processes will escape down ISM field lines in
competition with perpendicular scattering outwards. Wind sweeping is slow.
If the experimental gradient measures the trapping region as about 1 AU thick,
we simply estimate the escape time down the field as equal to the perpendicular
diffusion time over 1 AU. For motion over $d$, from
\begin{equation}
d=2(Kt_{esc})^{0.5}
\end{equation}
we find $t_{esc}=3\rightarrow8\times10^{4}s$. 
In this time 6 MeV particles can go $5\rightarrow8$ AU
along the ISM field. We would require the ISM field to be at a significant 
angle to the Heliopause for such a movement to be consistent with escape
from the postulated turbulent layer.\\
In conclusion, from the discussion of possible boundary models, a moving
boundary driven by changes in solar wind pressure is more likely than
a stationary boundary scan by Voyager 1. The particle results available also
seem compatible with the existence of turbulent regions
at or beyond the Heliopause,
of dimension 1 AU wide and up to 10 AU
 along the boundary, capable of
partially trapping an escaping flux of anomalous component particles 
which are mainly constrained to motion parallel to and inside the Heliopause.   

\section{ACKNOWLEDGMENTS}

W R Webber thanks his Voyager 1 colleagues from the CRS instrument,
 especially the late Frank McDonald and Project PI Ed Stone.
                    
\section{REFERENCES}
\noindent
 Borovikov S. N., Pogorelov N. V., Burlaga L. F., Richardson \\
\indent 
J. D., 2011, ApJ, 728 L21\\
\noindent
Burlaga L. F., Ness N. F., Acuna M. H., Richardson J. D.,\\
\indent
 Stone E. C., McDonald F. B.,2009 ApJ, 692, 1135 \\    
\noindent
Burlaga l. F., Ness N. F., 2012, ApJ., 749, 13\\ 
Burlaga l. F., Ness N. F., 2010, ApJ., 725, 1306\\ 
\noindent
Burlaga l. F., Ness N. F., 2012, ApJ., 749, 13\\ 
\noindent
Burlaga l. F., Ness N. F., Stone E. C., 2013, \\
\indent
Science, 341, 150\\
\noindent
Gurnett D. A., Kurth W. S., Cairns I. H., Mitchell J., 2006\\
\indent
in Heerikhuisen J.,Florinski V., Zank G. P., Pogorelov\\
\indent
N. V., eds, Physics of the Inner Heliosheath. American\\
\indent
Inst. Phys. p.129 \\
\noindent
\noindent
Jokipii J. R., Giacaloni J., Kota J., 2004 ApJ 611, L141\\
\noindent
Krimigis S. M.,
Roelof E. C., Decker R. B., Hill M. E., \\
\indent
2011, Nature 474, 359 \\
Krimigis S. M., Decker R. B., Roelof E. C.\\ 
\indent
Hill M. E., Armstrong T. P.,Gloeckler G.,\\
\indent
Hamilton D. C., Lanzerotti, L. J., 2013\\ 
\indent
Science, 341, 144\\   
\noindent
McComas D.J., Alexashov D. A., Browski M., Fahr H.,\\
\indent
 Heerikhuisen J.,Izmodenov V., 
Lee M.A., Mobius\\
\indent
M. A., Pogorelov N., Schwadron N.A., Zank G.P., 2012\\
\indent
Science, 336, 1291\\  
Nkosi,G. S., Potgieter  M. S., Webber W. R., 2011,\\
\indent
 Adv.in Space Res., 48, 1480\\
\noindent
Opher M., Stone E. C.,
Liewer P.C., 2006 ApJ. 640 L71 \\
\noindent
Opher, M., 
Drake J. F., Velli M., Decker R.B., Tooth G.,\\
\indent
2012, ApJ, 751, 80\\
Pei C., Bieber J. W., Breech B., Burger R. A.,\\
\indent
Clem J., Matthaeus W. H., 2010, \\
\indent
J. Geophys. Res.,115,A03103 \\
\noindent 
Potgieter M. S., Le Roux J. A., 1989, A \& A, 209, 406 \\ 
\noindent
 Potgieter, M. 
S., 2008, Advances in Space Sci. 41, 245\\
\noindent
Quenby J. J., Lockwood J. A., Webber W. R., 1990,\\
\indent
 ApJ, 365, 365\\
Scherer K., Fichtner H., Strauss R.D.,\\ 
\indent
Ferreira S. E., Busning L., Potgieter M.S.,\\
\indent
2008, Ap J., 680, L105 \\
\noindent
Stone E. C.,
Vogt R. E., McDonald F. B., Teegarden B. J.,\\
\indent
 Trainor J. F., Jokippi, J. R., Webber W. R., 1977,\\
\indent
 Space Science Reviews, 21, 355\\    
\noindent
Stone E, C., 
Cummings A. C., McDonald F. B., Heikkila\\
\indent
B., Lal N., Webber W. R., 2005,
Science, 23, 2017  \\
Strauss R.D., Potgieter M.S., Ferreira S. E. S.,\\ 
\indent
Fichtner H., Scherer K., 2013,\\
\indent
 ApJ. Lett. 76 L18\\
\noindent
\noindent
Webber W.R., McDonald F.B., 2013a,\\ 
\indent
Geophys.Res.Lett., 40, 1665\\
\noindent
Webber W.R., Higbie P. R., McDonald F.B., 2013b,\\
\indent
 arxiv..org ;1308.1895\\

\end{document}